\documentclass[10pt,conference]{IEEEtran}
\IEEEoverridecommandlockouts
\usepackage{cite}
\usepackage{amsmath,amssymb,amsfonts}
\usepackage{graphicx}
\usepackage{textcomp}
\usepackage{xcolor}
\usepackage{colortbl}
\usepackage{booktabs}
\usepackage{multirow}
\usepackage{makecell}
\def\BibTeX{{\rm B\kern-.05em{\sc i\kern-.025em b}\kern-.08em
    T\kern-.1667em\lower.7ex\hbox{E}\kern-.125emX}}
\begin{document}

\title{SoK: Adversarial Robustness of the Variational Quantum Eigensolver via Red-Teaming\\}

\author{
\IEEEauthorblockN{
Ahmed Azaz Humdoon\IEEEauthorrefmark{1},
Cheng Chu\IEEEauthorrefmark{2},
Lei Jiang\IEEEauthorrefmark{3},
Qian Lou\IEEEauthorrefmark{1},
Mengxin Zheng\IEEEauthorrefmark{1}
}
\IEEEauthorblockA{
\IEEEauthorrefmark{1}University of Central Florida \quad
\IEEEauthorrefmark{2}North Carolina State University \quad
\IEEEauthorrefmark{3}Indiana University Bloomington\\
\{ahmedazaz.humdoon, qian.lou, mengxin.zheng\}@ucf.edu \quad
cchu8@ncsu.edu \quad jiang60@iu.edu}
}

\maketitle

\begin{abstract}
The Variational Quantum Eigensolver (VQE) is a leading algorithm for estimating molecular ground-state energies on near-term quantum hardware, with applications spanning quantum chemistry, materials science, and drug discovery. As VQE workloads are increasingly deployed through cloud-based ``VQE-as-a-service'' pipelines, they become exposed to adversaries such as compromised service components, malicious co-tenants, or insiders in the transpilation stack, any of which can corrupt results before they reach the user. A range of attacks on variational quantum circuits has been proposed, but each has been studied in isolation: some on quantum classifiers with accuracy-based metrics, others on variational quantum algorithms with energy-error metrics. This lack of a common evaluation setup makes their relative severity difficult to compare and leaves the security of VQE poorly characterized. In this work, we present \textbf{VQE-AdvBench}, the first unified red-teaming benchmark for the Variational Quantum Eigensolver, systematizing these attacks under a single evaluation protocol to rigorously assess VQE's adversarial robustness. We organize attacks along a black-, gray-, and white-box access taxonomy, and evaluate seven representative attack scenarios---the QTrojan circuit backdoor, the QDoor parameter backdoor, parameter-space adaptations of FGSM and PGD, and three QNBAD noise-induced variants---over a fixed molecule--ansatz--backend--metric configuration, on H$_2$ and H$_3^+$ across five noise-calibrated IBM backends. Our results reveal a clear severity ordering: noise-induced attacks that manipulate the Zero-Noise Extrapolation (ZNE) pipeline are the most damaging (up to 8.84$\times$ error amplification), followed by the QTrojan circuit-level backdoor (7.52$\times$), while the QDoor parameter-level backdoor is the least effective, yielding only marginal amplification (up to 1.37$\times$). 
\end{abstract}


\section{Introduction}

The Variational Quantum Eigensolver (VQE)~\cite{peruzzo2014variational} is one of the most prominent variational quantum algorithms (VQAs)~\cite{cerezo2021variational}. It is a hybrid quantum-classical algorithm: a parameterized circuit $U(\boldsymbol{\theta})$ prepares a trial state whose energy $\langle\psi(\boldsymbol{\theta})|H|\psi(\boldsymbol{\theta})\rangle$ is minimized by a classical optimizer over the parameter vector $\boldsymbol{\theta}$, producing an upper bound on the ground-state energy of a Hamiltonian $H$. VQE is widely used in quantum chemistry to compute molecular energies and electronic properties~\cite{tilly2022variational}, and has been applied across domains including drug discovery~\cite{blunt2022perspective}, materials science~\cite{lordi2021advances}, and chemical engineering~\cite{cao2019quantum}.

VQE workloads are increasingly deployed through a cloud-based \emph{VQE-as-a-service} model, where a user submits a Hamiltonian and ansatz to a remote service and receives expectation values or optimized parameters in return. In this setting, the service-side pipeline has full visibility into the circuit, variational parameters, and target backend. A compromised service component, a malicious co-tenant exploiting shared-hardware effects, or an insider in the transpilation pipeline could use this access to introduce circuit-level backdoors, perturb parameter values, or shape noise-dependent errors before results reach the user. Because optimized parameter vectors are often reused across downstream workflows, a poisoned solution can propagate silently. The victim then runs what appears to be a valid VQE workflow while receiving a corrupted ground-state estimate. As reliance on VQE grows for high-stakes decisions in chemistry, materials science, and drug discovery, its security becomes a critical concern. For example, in drug discovery, inaccurate quantum simulations can identify ineffective compounds as promising candidates, leading to wasted resources and possible risks to patient safety~\cite{lazic2021quantifying}. Small errors of 1--2 kcal/mol can underestimate impact sensitivity, causing hazardous compounds to be misclassified and pass safety checks, potentially leading to catastrophic failure~\cite{Duarte2023WhichMP}. These examples highlight the need to ensure the robustness and security of VQE systems.

A range of attacks on parameterized quantum circuits has been proposed---circuit-level backdoors, parameter-space poisoning, gradient-based perturbations, and noise-induced manipulation---but each has been studied in isolation, on task-specific datasets and metrics. This heterogeneity leaves their relative severity resistant to direct comparison and obscures which threats matter most for VQE. Prior systematization is limited: the SoK of Nowmi \emph{et al.}~\cite{nowmi2025critical} evaluates only quantum machine learning (QML) \emph{classifiers}, but does not cover variational algorithms such as VQE and does not rank attacks by severity for VQA objectives. Red-teaming, which applies structured adversarial stress testing under a common protocol, offers a principled way to close this gap. It is the basis for standardized ML-safety benchmarks such as HarmBench~\cite{mazeika2024harmbench} and RobustBench~\cite{croce2020robustbench}, yet no comparable benchmark exists for variational quantum algorithms.

In this work we present \textbf{VQE-AdvBench}, the first unified red-teaming benchmark that systematizes these threats under a single evaluation protocol. We organize attacks along an access-level taxonomy (black-, gray-, white-box) and evaluate seven representative attack scenarios---the QTrojan circuit backdoor~\cite{chu2023qtrojan}, the QDoor parameter backdoor~\cite{chu2023qdoor}, parameter-space adaptations of FGSM~\cite{goodfellow2014explaining} and PGD~\cite{madry2017towards}, and three QNBAD noise-induced variants~\cite{chuqnbad} (FreeDrift, MimicSlope, SilentShift)---over fixed molecule--ansatz--backend--metric configurations, on H$_2$ and H$_3^+$ across five noise-calibrated IBM backends. Our evaluation produces the first head-to-head severity ranking of these attacks under a common protocol, revealing which categories of attack are most damaging to VQE.

\section{Background}
\label{sec:background}
This section introduces essential background concepts for this work. We first describe the Variational Quantum Eigensolver (VQE) and then Zero-Noise Extrapolation (ZNE). ZNE is an error-mitigation step that the QNBAD variants attack.

\medskip
\noindent\textbf{Variational Quantum Eigensolver.}
VQE estimates the ground-state energy of a Hamiltonian $H$ using the variational principle: for any normalized state $|\psi(\boldsymbol{\theta})\rangle$, the expectation value $E(\boldsymbol{\theta}) = \langle\psi(\boldsymbol{\theta})|H|\psi(\boldsymbol{\theta})\rangle$ is bounded below by the true ground-state energy $E_0$. A parameterized circuit $U(\boldsymbol{\theta})$ prepares the trial state $|\psi(\boldsymbol{\theta})\rangle = U(\boldsymbol{\theta})|\psi_0\rangle$. A classical optimizer (e.g., COBYLA or Adam) iteratively updates $\boldsymbol{\theta}$ to minimize $E(\boldsymbol{\theta})$, yielding an upper bound on $E_0$~\cite{cerezo2021variational}. 

\medskip
\noindent\textbf{Zero-Noise Extrapolation.}
ZNE~\cite{temme2017error} is a widely used error-mitigation strategy for NISQ devices. It deliberately amplifies circuit noise by integer scaling factors $\lambda_1,\dots,\lambda_K$, measures the noisy observable at each level, and extrapolates the resulting series back to the zero-noise limit ($\lambda\rightarrow 0$). Because ZNE reconstructs the ideal value from a fitted noise trajectory, an adversary who manipulates that trajectory can corrupt the extrapolated result.

\section{Threat Model and Attack Taxonomy}
\label{sec:taxonomy}

We organize attacks by the adversary's access level: black-box (query-only), gray-box (partial visibility into intermediate artifacts such as transpiled circuits), and white-box (full access to parameters, architecture, and hardware). All attacks we evaluate assume a compromised or untrusted cloud VQE service and aim to corrupt the ground-state estimate. Attacks that require a labeled dataset or an input distribution (e.g., label-flipping poisoning, query-based evasion against classifiers) do not transfer to VQE, which has no labels or classical input, and are excluded. We summarize the seven evaluated attacks below.

\medskip
\noindent\textbf{Circuit-level backdoor (QTrojan, gray-box).} A semi-privileged adversary inserts concealed pre- and post-encoding $R_X$/$R_Y$ layers, disguised as calibration procedures, around the state-preparation stage~\cite{chu2023qtrojan}. A server-specific configuration file acts as the trigger: when disabled, the circuit behaves like its clean counterpart; when enabled, it amplifies error on the target backend, corrupting the ground-state estimate and compromising integrity.

\medskip
\noindent\textbf{Parameter-level backdoor (QDoor, white-box).} An insider with full circuit access embeds malicious behavior into the trained parameters via approximate-synthesis adversarial training~\cite{chu2023qdoor}, so that the backdoor activates once the synthesized circuit is executed on hardware.

\medskip
\noindent\textbf{Gradient-based evasion (FGSM/PGD, white-box).} Classical evasion attacks like FGSM~\cite{goodfellow2014explaining} and PGD~\cite{madry2017towards} perturb an input to cross a decision boundary. VQE has no classical input; the only quantity the user trains and deploys is the parameter vector $\boldsymbol{\theta}^*$. We therefore adapt both attacks to perturb $\boldsymbol{\theta}^*$ directly. FGSM is a one-shot attack: it takes a single step of size $\varepsilon$ that pushes the parameters in the direction that increases the energy error. PGD is the iterative version: it takes several smaller steps in that same direction, keeping the total change within a fixed budget $\varepsilon$.

\medskip
\noindent\textbf{Noise-induced backdoor (QNBAD, white-box).} The adversary delivers parameters that look benign in noiseless simulation but shape how noisy expectation values behave across ZNE scaling levels, so that extrapolation returns a value far from the true energy ~\cite{chuqnbad}. Its three variants FreeDrift (QF), MimicSlope (QM), and SilentShift (QS) differ in how aggressively they distort the noise trajectory: QF maximally changes sampled values across levels, QM uniformly shifts the trajectory, and QS keeps low-noise values unchanged while reducing high-noise values.

\section{Experimental Methodology}
\label{sec:setup}

We deliberately scope this first benchmark to a controlled setting: small, well-characterized molecules and a fixed ansatz family. This controlled scope lets us compare attack severity cleanly across a common molecule, ansatz, backend, and metric configuration.

\medskip
\noindent\textbf{Dataset.} We use the PennyLane Molecules dataset~\cite{azad2023pennylane}, selecting H$_2$ and H$_3^+$, which correspond to 4-qubit and 6-qubit Hamiltonians respectively. The fermionic Hamiltonians are mapped to qubit Hamiltonians using the Jordan-Wigner transformation~\cite{1928ZPhy...47..631J}.

\medskip
\noindent\textbf{Ansatz.} We use the Hardware-Efficient Ansatz \texttt{efficient\_su2} from Qiskit~\cite{javadi2024quantum} with linear entanglement: $\mathrm{reps}=2$ (24 parameters) for H$_2$ and $\mathrm{reps}=3$ (48 parameters) for H$_3^+$.

\medskip
\noindent\textbf{Noisy Backends.} We evaluate each attack on five IBM Quantum fake backends from \texttt{qiskit\_ibm\_runtime.fake\_provider}: \texttt{fake\_cairo} (CAI), \texttt{fake\_montreal} (MON), \texttt{fake\_guadalupe} (GUA), \texttt{fake\_almaden} (ALM), and \texttt{fake\_auckland} (AUC). These incorporate realistic noise models calibrated from real hardware and span a range of device generations and qubit counts.

\medskip
\noindent\textbf{ZNE Setting.} ZNE is configured with Mitiq~\cite{larose2022mitiq}, scaling noise by factors $T=\{1,\dots,6\}$ and fitting a degree-2 polynomial to the sampled expectation values.

\medskip
\noindent\textbf{Evaluation Metric.} We use the absolute error between the adversarial and ideal energy estimates, $E_{\text{abs}} = |E_{\text{adv}} - f_{\text{ideal}}|$, where $f_{\text{ideal}}$ is the noise-free expectation value. A higher $E_{\text{abs}}$ (equivalently, a higher amplification relative to the clean baseline) indicates a more successful attack.

\section{Results}
\label{sec:results}

We report attack effectiveness on VQE for H$_2$ and H$_3^+$ across the five backends. Each cell reports $E_{\text{abs}}$ with the amplification factor relative to the clean baseline in parentheses; the GEO-Mean column is the geometric mean across the two molecules. Circuit- and parameter-space attacks (Table~\ref{tab:res-master}) are compared against a single-noisy clean baseline; the QNBAD noise-induced attacks (Table~\ref{tab:res-qnbad}) target ZNE and are compared against a ZNE-extrapolated clean baseline so that attacked and clean estimates share the same pipeline.

\subsection{Circuit- and Parameter-Space Attacks}
\label{sec:res-master}

For QTrojan, QDoor, FGSM, and PGD we used the Adam optimizer (300 iterations, learning rate 0.1); FGSM and PGD are reported at $\varepsilon = 0.25$~rad. Table~\ref{tab:res-master} shows QTrojan is the strongest in this group, followed by PGD, FGSM, and QDoor. QTrojan amplification reaches up to 9.88$\times$ on H$_2$ and 5.72$\times$ on H$_3^+$, peaking at a 7.52$\times$ geometric mean on MON. PGD consistently outperforms FGSM under the same perturbation budget (geometric means 3.32--6.15$\times$ vs.\ 2.07--3.38$\times$), confirming that the iterative, projected update is more damaging than the single-step one. QDoor stays near the clean baseline (0.94--1.37$\times$), likely because its approximate-synthesis trigger has little room to inject error in these small, shallow circuits. QTrojan, FGSM, and PGD all peak on MON and bottom out on GUA.

\begin{table}[!t]
\centering
\caption{Effectiveness of QTrojan, QDoor, FGSM, and PGD against VQE on H$_2$ and H$_3^+$ across five IBM backends.}
\label{tab:res-master}
\renewcommand{\arraystretch}{1.1}
\setlength{\tabcolsep}{4pt}
\footnotesize
\begin{tabular}{|c|l|c|c|c|}
\hline
\multirow{2}{*}{\textbf{Devices}} & \multirow{2}{*}{\textbf{Schemes}} & \multicolumn{3}{c|}{\textbf{Tasks (absolute error ($\times$ rel.\ to clean))}} \\
\cline{3-5}
 & & \textbf{VQE-H$_2$} & \textbf{VQE-H$_3^+$} & \textbf{GEO-Mean} \\
\hline
\multirow{5}{*}{CAI}
  & Clean   & 0.0770 (1.0$\times$) & 0.1980 (1.0$\times$) & 0.1235 (1.0$\times$) \\
\cline{2-5}
  & QTrojan & 0.4820 (6.26$\times$) & 0.6364 (3.21$\times$) & 0.5538 (4.48$\times$) \\
\cline{2-5}
  & QDoor   & 0.0905 (1.17$\times$) & 0.1545 (0.78$\times$) & 0.1182 (0.96$\times$) \\
\cline{2-5}
  & FGSM    & 0.2326 (3.02$\times$) & 0.3490 (1.76$\times$) & 0.2849 (2.31$\times$) \\
\cline{2-5}
  & PGD     & 0.3036 (3.94$\times$) & 0.6980 (3.53$\times$) & 0.4603 (3.73$\times$) \\
\hline
\multirow{5}{*}{MON}
  & Clean   & 0.0471 (1.0$\times$) & 0.1025 (1.0$\times$) & 0.0695 (1.0$\times$) \\
\cline{2-5}
  & QTrojan & 0.4655 (9.88$\times$) & 0.5862 (5.72$\times$) & 0.5224 (7.52$\times$) \\
\cline{2-5}
  & QDoor   & 0.0491 (1.04$\times$) & 0.1090 (1.06$\times$) & 0.0732 (1.05$\times$) \\
\cline{2-5}
  & FGSM    & 0.2067 (4.39$\times$) & 0.2671 (2.61$\times$) & 0.2350 (3.38$\times$) \\
\cline{2-5}
  & PGD     & 0.2807 (5.96$\times$) & 0.6505 (6.35$\times$) & 0.4273 (6.15$\times$) \\
\hline
\multirow{5}{*}{GUA}
  & Clean   & 0.1085 (1.0$\times$) & 0.1883 (1.0$\times$) & 0.1429 (1.0$\times$) \\
\cline{2-5}
  & QTrojan & 0.5002 (4.61$\times$) & 0.6336 (3.36$\times$) & 0.5630 (3.94$\times$) \\
\cline{2-5}
  & QDoor   & 0.1045 (0.96$\times$) & 0.1724 (0.92$\times$) & 0.1342 (0.94$\times$) \\
\cline{2-5}
  & FGSM    & 0.2579 (2.38$\times$) & 0.3390 (1.80$\times$) & 0.2957 (2.07$\times$) \\
\cline{2-5}
  & PGD     & 0.3270 (3.02$\times$) & 0.6892 (3.66$\times$) & 0.4747 (3.32$\times$) \\
\hline
\multirow{5}{*}{ALM}
  & Clean   & 0.0568 (1.0$\times$) & 0.2140 (1.0$\times$) & 0.1103 (1.0$\times$) \\
\cline{2-5}
  & QTrojan & 0.4706 (8.28$\times$) & 0.6424 (3.00$\times$) & 0.5498 (4.98$\times$) \\
\cline{2-5}
  & QDoor   & 0.0900 (1.58$\times$) & 0.2550 (1.19$\times$) & 0.1515 (1.37$\times$) \\
\cline{2-5}
  & FGSM    & 0.2157 (3.80$\times$) &  0.3702 (1.73$\times$) & 0.2826 (2.56$\times$) \\
\cline{2-5}
  & PGD     & 0.2871 (5.05$\times$) & 0.7027 (3.28$\times$) & 0.4492 (4.07$\times$) \\
\hline
\multirow{5}{*}{AUC}
  & Clean   & 0.0602 (1.0$\times$) & 0.1283 (1.0$\times$) & 0.0879 (1.0$\times$) \\
\cline{2-5}
  & QTrojan & 0.4725 (7.85$\times$) & 0.5998 (4.67$\times$) & 0.5324 (6.05$\times$) \\
\cline{2-5}
  & QDoor   & 0.0641 (1.07$\times$) & 0.1225 (0.95$\times$) & 0.0886 (1.01$\times$) \\
\cline{2-5}
  & FGSM    & 0.2179 (3.62$\times$) & 0.2894 (2.26$\times$) & 0.2511 (2.86$\times$) \\
\cline{2-5}
  & PGD     & 0.2912 (4.84$\times$) & 0.6630 (5.17$\times$) & 0.4394 (5.00$\times$) \\
\hline
\end{tabular}
\end{table}

\subsection{Noise-Induced Attacks}
\label{sec:res-qnbad}

For the three QNBAD variants we used SPSA with Adam via PennyLane (300 steps, learning rate 0.1). Table~\ref{tab:res-qnbad} shows all three can substantially increase the ZNE-extrapolated error. QF is the strongest overall, with a peak geometric mean of 8.84$\times$ on CAI (per-molecule peaks of 11.35$\times$ for H$_2$ and 12.48$\times$ for H$_3^+$). QM is also highly effective, with a peak geometric mean amplification of 8.35$\times$ on CAI and peak per-molecule amplifications of 6.55$\times$ for H$_2$ on ALM and 12.30$\times$ for H$_3^+$ on CAI. QS is consistently weaker than QF and QM but still not negligible, with a peak geometric mean amplification of 4.52$\times$ on CAI and per-molecule peaks of 4.32$\times$ for H$_2$ and 4.73$\times$ for H$_3^+$, both on CAI. These results show that attacking the ZNE pipeline can be damaging even when the executed circuit is not directly modified, since the attack changes how the noisy expectation values behave during extrapolation.

\begin{table}[!t]
\centering
\caption{Effectiveness of three QNBAD variants against VQE on H$_2$ and H$_3^+$ across five backends. QF = FreeDrift, QM = MimicSlope, QS = SilentShift.}
\label{tab:res-qnbad}
\renewcommand{\arraystretch}{1.1}
\setlength{\tabcolsep}{4pt}
\footnotesize
\begin{tabular}{|c|l|c|c|c|}
\hline
\multirow{2}{*}{\textbf{Devices}} & \multirow{2}{*}{\textbf{Schemes}} & \multicolumn{3}{c|}{\textbf{Tasks (absolute error ($\times$ rel.\ to clean))}} \\
\cline{3-5}
 & & \textbf{VQE-H$_2$} & \textbf{VQE-H$_3^+$} & \textbf{GEO-Mean} \\
\hline
\multirow{4}{*}{CAI}
  & Clean & 0.0166 (1.0$\times$) & 0.0231 (1.0$\times$) & 0.0196 (1.0$\times$) \\
\cline{2-5}
  & QF    & 0.1038 (6.26$\times$) & 0.2889 (12.48$\times$) & 0.1732 (8.84$\times$) \\
\cline{2-5}
  & QM    & 0.0939 (5.67$\times$) & 0.2846 (12.30$\times$) & 0.1635 (8.35$\times$) \\
\cline{2-5}
  & QS    & 0.0716 (4.32$\times$) & 0.1095 (4.73$\times$) & 0.0885 (4.52$\times$) \\
\hline
\multirow{4}{*}{MON}
  & Clean & 0.0471 (1.0$\times$) & 0.1045 (1.0$\times$) & 0.0701 (1.0$\times$) \\
\cline{2-5}
  & QF    & 0.5346 (11.35$\times$) & 0.2782 (2.66$\times$) & 0.3856 (5.49$\times$) \\
\cline{2-5}
  & QM    & 0.0981 (2.08$\times$) & 0.3245 (3.11$\times$) & 0.1784 (2.54$\times$) \\
\cline{2-5}
  & QS    & 0.0633 (1.34$\times$) & 0.2784 (2.67$\times$) & 0.1328 (1.89$\times$) \\
\hline
\multirow{4}{*}{GUA}
  & Clean & 0.1085 (1.0$\times$) & 0.1866 (1.0$\times$) & 0.1423 (1.0$\times$) \\
\cline{2-5}
  & QF    & 0.1922 (1.77$\times$) & 0.2418 (1.30$\times$) & 0.2156 (1.52$\times$) \\
\cline{2-5}
  & QM    & 0.1818 (1.68$\times$) & 0.4304 (2.31$\times$) & 0.2797 (1.97$\times$) \\
\cline{2-5}
  & QS    & 0.1040 (0.96$\times$) & 0.4469 (2.39$\times$) & 0.2156 (1.51$\times$) \\
\hline
\multirow{4}{*}{ALM}
  & Clean & 0.0568 (1.0$\times$) & 0.2140 (1.0$\times$) & 0.1103 (1.0$\times$) \\
\cline{2-5}
  & QF    & 0.2004 (3.53$\times$) & 0.4982 (2.33$\times$) & 0.3160 (2.86$\times$) \\
\cline{2-5}
  & QM    & 0.3719 (6.55$\times$) & 0.4451 (2.08$\times$) & 0.4069 (3.69$\times$) \\
\cline{2-5}
  & QS    & 0.1206 (2.12$\times$) & 0.2572 (1.20$\times$) & 0.1761 (1.60$\times$) \\
\hline
\multirow{4}{*}{AUC}
  & Clean & 0.0602 (1.0$\times$) & 0.1283 (1.0$\times$) & 0.0879 (1.0$\times$) \\
\cline{2-5}
  & QF    & 0.1313 (2.18$\times$) & 0.3656 (2.85$\times$) & 0.2191 (2.49$\times$) \\
\cline{2-5}
  & QM    & 0.1697 (2.82$\times$) & 0.8302 (6.47$\times$) & 0.3753 (4.27$\times$) \\
\cline{2-5}
  & QS    & 0.0957 (1.59$\times$) & 0.4025 (3.14$\times$) & 0.1963 (2.23$\times$) \\
\hline
\end{tabular}
\end{table}

\section{Defenses}
\label{sec:defenses}

We outline defenses matched to the evaluated attacks, leaving their empirical evaluation to future work. Since these attacks target different stages of the VQE pipeline, no single countermeasure defends against all of them. Against encoding- and circuit-level backdoors (QTrojan), quantum-aware validation and classical--quantum consistency checks can detect tampering at the state-preparation stage. Against noise-induced backdoors (QNBAD), classical backdoor defenses such as fine-pruning~\cite{liu2018fine} may help but need careful calibration in the VQA setting, while varying the noise model or applying randomized compiling and dynamical decoupling~\cite{wallman2016noise} disrupts the device-specific triggers QNBAD relies on. Against white-box parameter-space attacks, partitioning the circuit across isolated backends, as in QuMoS~\cite{wang2023qumos}, limits an adversary's access, since no single provider controls the full workflow.

\section{Discussion and Future Work}
\label{sec:discussion}

Several attack classes are deliberately out of scope. \emph{Pulse-level attacks} require hardware- and physics-level access below the circuit abstraction, whereas we operate at the circuit and parameter level exposed by Qiskit and PennyLane. \emph{Black-box query-only attacks} (e.g., label flipping) assume a labeled dataset and input distribution that VQE lacks. \emph{Compiler-internal attacks} assume an untrusted compiler, whereas we assume the compiler is trusted.

The benchmark can be extended in a few directions. The unified protocol could be applied to other VQAs such as QAOA and VQD. Extending to larger molecules such as H$_4$ ($\sim$8 qubits), LiH ($\sim$12 qubits), and BeH$_2$ ($\sim$14 qubits) would test how attack amplification scales with circuit depth, parameter count, and hardware noise. We leave this larger-scale study to future work.

\section{Conclusion}
\label{sec:conclusion}

In this work we introduced VQE-AdvBench, a red-teaming benchmark that evaluates seven attack scenarios spanning circuit-level, parameter-level, gradient-based, and noise-induced threats against the Variational Quantum Eigensolver, under a unified protocol fixing the molecule, ansatz, backend, and metric. Our central finding is that attack severity varies sharply across these classes: noise-induced attacks on the ZNE pipeline are the most damaging, followed by the QTrojan circuit-level backdoor and our FGSM/PGD adaptations, while the QDoor parameter-level backdoor is the weakest on these shallow circuits. We hope that VQE-AdvBench encourages the community to adopt standardized red-teaming as a common practice for evaluating the security of variational quantum algorithms. 

\bibliographystyle{IEEEtran}
\bibliography{references}

\end{document}